\documentclass[%
reprint,
superscriptaddress,
showpacs,
amsmath,amssymb,
prc,
floatfix ]%
{revtex4-1}

\usepackage{color}
\usepackage{CJK}

\usepackage{graphicx}
\usepackage{dcolumn}
\usepackage{bm}
\usepackage{hyperref}

\allowdisplaybreaks

\begin{document}

\begin{CJK*}{UTF8}{}
\title{Time-dependent generator coordinate method study of mass-asymmetric fission of actinides}
\CJKfamily{gbsn}
\author{Jie Zhao (赵杰)}%
\affiliation{Microsystem and Terahertz Research Center and Insititute of Electronic Engineering, 
	China Academy of Engineering Physics, Chengdu 610200, Sichuan, China}
\CJKfamily{gbsn}
\author{Jian Xiang (向剑)}%
\affiliation{Department of Physics and Electronic Science, Qiannan Normal University for Nationalities, Duyun, 558000, China}
\CJKfamily{gbsn}
\author{Zhi-Pan Li (李志攀)}%
\affiliation{School of Physical Science and Technology, Southwest University, Chongqing 400715, China}
\author{Tamara Nik\v{s}i\'c}%
\affiliation{Physics Department, Faculty of Science, University of Zagreb, Bijeni\v{c}ka Cesta 32,
        	      Zagreb 10000, Croatia}             
\author{Dario Vretenar}%
\affiliation{Physics Department, Faculty of Science, University of Zagreb, Bijeni\v{c}ka Cesta 32,
              Zagreb 10000, Croatia}
\CJKfamily{gbsn}              
\author{Shan-Gui Zhou (周善贵)}%
 \affiliation{CAS Key Laboratory of Theoretical Physics, Institute of Theoretical Physics, Chinese Academy of Sciences, Beijing 100190, China}
 \affiliation{School of Physical Sciences, University of Chinese Academy of Sciences, Beijing 100049, China}
 \affiliation{Center of Theoretical Nuclear Physics, National Laboratory of Heavy Ion Accelerator, Lanzhou 730000, China}
 \affiliation{Synergetic Innovation Center for Quantum Effects and Application, Hunan Normal University, Changsha 410081, China}

\date{\today}

\begin{abstract}
Low-energy positive and negative parity collective states in the equilibrium minimum, and the dynamics of 
 induced fission of actinide nuclei are investigated in a unified theoretical framework based on the generator coordinate 
 method (GCM) with the Gaussian overlap approximation (GOA).
The collective potential and inertia tensor, both at zero and finite temperature,
are computed using the self-consistent multidimensionally constrained relativistic mean field (MDC-RMF) model, 
based on the energy density functional DD-PC1. Pairing correlations are treated in the BCS approximation with a 
separable pairing force of finite range.  A collective quadrupole-octupole Hamiltonian 
characterized by zero-temperature axially-symmetric deformation energy surface and perturbative 
cranking inertia tensor, is used to model the low-lying excitation spectrum. 
The fission fragment charge distributions are obtained by propagating the initial collective states 
in time with the time-dependent GCM+GOA that uses the same quadrupole-octupole Hamiltonian, 
but with the collective potential and inertia tensor computed at finite temperature. The illustrative charge yields of 
$^{228}$Th, $^{234}$U, $^{240}$Pu, $^{244}$Cm, and $^{250}$Cf  are in very good agreement with experiment, 
and the predicted mass asymmetry is consistent with the result of a recent microscopic study that has attributed 
the distribution (peak) of the heavier-fragment nuclei to shell-stabilized octupole deformations.
\end{abstract}

\maketitle

\end{CJK*}

\bigskip

\section{Introduction~\label{sec:Introduction}}

Quantum shell effects determine the full spectrum of nuclear structure phenomena, from the formation of clusters in 
light nuclei to the stability of superheavy systems. In particular, the mass and charge distribution of fission fragments 
is governed by the shell structure of the effective nuclear potential. 
Fission intrinsically presents a dynamical process in which a quasi-static initial nuclear state evolves with time towards 
a two-fragment final configuration \cite{Schunck2016_RPP79-116301}. In a recent study \cite{Scamps2018_Nature564-382}, 
based on the concept of time-dependent (TD) density functional theory (DFT) \cite{Negele1978_PRC17-1098}, it has 
been shown that the final mass asymmetry of the fragments  in the fission of heavy (actinide) nuclei is, to a large extent,
determined by the extra stability of heavier-fragment nuclei with charge number between $Z=52$ and $Z=56$, 
characterized by pronounced octupole deformation. 

The fully microscopic and nonadiabatic TDDFT describes the dynamics of the fission process starting from an adiabatic 
configuration just beyond the saddle, and ending with separate fragments. It has been shown that many collective degrees 
of freedom are excited in the fission process, and that one-body dissipation plays an important role \cite{Bulgac2016_PRL116-122504}. Vibrational modes of post-scission fragments have also been investigated in this 
framework \cite{Simenel2014_PRC89-031601}.
Physical observables such as the most probable charge, mass, and total kinetic energy yields 
can be extracted from the TDDFT calculations. However, a realistic TDDFT description of the entire fission process, 
including the first phase from the ground-state potential well to beyond the fission barrier, is still not possible. 
 Even though the stochastic extension of the standard TDDFT provides a possible solution~\cite{Tanimura2017_PRL118-152501},
applications are still limited because of its computational complexity.
It is also well known that the quantum tunneling process cannot be described 
with TDDFT due to its semiclassical nature~\cite{Negele1982_RMP54-913}.

An alternative microscopic approach capable of predicting both the low-energy collective excitation spectra 
in the deformed equilibrium minimum and the fission fragment distribution,
is the generator coordinate method (GCM)~\cite{Niksic2009_PRC79-034303,Prochniak2009_JPG36-123101,
Schunck2016_RPP79-116301,Berger1991_CPC63-365,Libert1999_PRC60-054301}.
In the Gaussian overlap approximation (GOA) the GCM Hill-Wheeler equation reduces to a local
Schr\"odinger-like equation in the space of collective coordinates.
For a specific choice of collective coordinates, the essential inputs are the potential and inertia tensor that 
can be computed microscopically in a self-consistent mean-field deformation-constrained calculation.
In the static case the low-lying excitation spectrum is obtained from the solution of the eigenvalue problem 
for the collective Hamiltonian. 
Starting from the initial state of the compound nucleus in the collective space, the 
adiabatic time evolution of the fissioning system is modelled with the time-dependent equation of the 
collective Hamiltonian, and the fission fragment distribution can be obtained by considering the flux of 
the probability current through the scission hyper-surface. TDGCM+GOA does not take into account 
non-adiabatic effects arising from the coupling between collective and intrinsic degrees of freedom.

Low-energy excitation spectra of actinide nuclei, characterized by pronounced octupole correlations, 
have successfully been described using a GCM+GOA Hamiltonian in the collective space 
of quadrupole and octupole deformations~\cite{Li2013_PLB726-866,Li2016_JPG43-024005,
Xia2017_PRC96-054303,Xu2017_CPC41-124107,Zhao2019_IJMPE27-1830007}.
The TDGCM+GOA, based on non-relativistic Skyrme  or Gogny functionals, has been applied to   
the analysis of fission dynamics of actinides in several studies \cite{Berger1991_CPC63-365,
Goutte2005_PRC71-024316,Younes2012_LLNL-TR-586678,
Regnier2018_CPC225-180,Regnier2017_EPJWC146-04043,Regnier2016_PRC93-054611,
Regnier2016_CPC200-350,Zdeb2017_PRC95-054608,Verriere2017_EPJWC146-04034,Regnier2019_PRC99-024611}. 
More recently, relativistic energy density functionals (EDF) \cite{Vretenar2005_PR409-101,Meng2006_PPNP57-470,Meng2016_WorldSci} 
have also been employed in the description of fission properties of heavy and superheavy 
nuclei~\cite{Zhou2016_PS91-063008,Burvenich2004_PRC69-014307,Blum1994_PLB323-262,Zhang2003_CPL20-1694,
Bender2003_RMP75-121,Lu2006_CPL23-2940,Li2010_PRC81-064321,Abusara2010_PRC82-044303,
Abusara2012_PRC85-024314,Lu2012_PRC85-011301R,Lu2014_PRC89-014323,Zhao2015_PRC91-014321,
Agbemava2017_PRC95-054324,Prassa2012_PRC86-024317,Zhao2015_PRC92-064315,Zhao2016_PRC93-044315}. 
The first study of fission dynamics that used the TDGCM+GOA based on a relativistic EDF 
was reported in Ref.~\cite{Tao2017_PRC96-024319}.

For the case of induced fission, one expects that the deformation energy surface of the fissioning nucleus 
and the collective inertia tensor will be modified as the internal excitation energy 
increases~\cite{Zhu2016_PRC94-024329,
Schunck2015_PRC91-034327,McDonnell2013_PRC87-054327,
McDonnell2014_PRC90-021302R,Pei2009_PRL102-192501,Martin2009_IJMPE18-861}.
Exploratory studies of finite-temperature (FT ) effects on induced fission yield distributions, based on  
semiclassical approaches, have been reported in Refs.~\cite{Ivanyuk2018_PRC97-054331,Randrup2013_PRC88-064606,Pasca2016_PLB760-800}.  
In the recent study of Ref.~\cite{Zhao2019_PRC99-014618} we have performed the first microscopic 
investigation of FT effect on induced fission yield distributions using the TDGCM+GOA collective model.
By considering the FT extension of nuclear density functional theory, 
a significant improvement is obtained for the predicted fission yields in comparison to data. The purpose of 
the present study is to show that the TDGCM+GOA based on nuclear energy density functionals can equally well 
be applied in the analysis of low-energy collective spectra in the equilibrium minimum and, when extended 
to finite temperature, to the description of the entire process of induced fission, using the same set of 
parameters of the microscopic EDF and pairing interaction. We will consider, in particular, actinide nuclei 
for which it has recently been shown that octupole correlations play a decisive role in the distribution of 
fission fragments \cite{Scamps2018_Nature564-382}. 
The theoretical framework and methods are introduced in Sec.~\ref{sec:model}.
The details of the calculation, the results for deformation energy surfaces, 
excitation spectra, as well as the charge yield distributions are described and discussed in Sec.~\ref{sec:results}.
Sec.~\ref{sec:summary} contains a summary of the principal results.

\section{\label{sec:model}Model}
The implementation of the TDGCM+GOA 
collective Hamiltonian method used in the present study 
is described in detail in Ref.~\cite{Li2016_JPG43-024005} (static aspects), 
and in Refs.~\cite{Tao2017_PRC96-024319,Zhao2019_PRC99-014618} (application to fission dynamics). 
For completeness here we include a brief outline of the model and 
discuss the basic approximations. 
Nuclear excitations characterized by quadrupole and octupole vibrational and rotational degrees 
of freedom can be described by considering quadrupole and octupole collective coordinates that 
specify the surface of a nucleus $R=R_0\left[1+\sum_\mu{\alpha_{2\mu}Y^*_{2\mu}+ \sum_\mu{\alpha_{3\mu}Y_{3\mu}^*} } \right]$ \cite{Li2016_JPG43-024005}.  
In addition, when axial symmetry is imposed, the collective coordinates can be parameterized in terms of 
two deformation parameters $\beta_{20}$ and $\beta_{30}$, and the Euler angles $\Omega$. 
In the GCM+GOA framework, after quantization the collective Hamiltonian reads
\begin{equation}
\hat{H}_{\rm coll}(\bm{q}) = -{\hbar^{2} \over 2\sqrt{w\mathcal{I}}} 
	\sum_{ij} {\partial \over \partial q_i}  \sqrt{\mathcal{I} \over w} B_{ij}(\bm{q}) {\partial \over \partial q_j}  
	+ {\hat{J}^2 \over 2\mathcal{I}} + V(\bm{q}),
\label{eq:Hcoll}
\end{equation} 
where $q_{i} \equiv \{\beta_{20}, \beta_{30}\}$, $V(\bm{q})$ denotes the collective potential, $B_{ij}(\bm{q})$ is the mass tensor.
$w=B_{22}B_{33} - B_{23}^{2}$, and $\mathcal{I}$ is the moment of inertia. 
The dynamics of the quadrupole-octupole collective Hamiltonian (QOCH) is governed by five functions of the intrinsic 
deformations $\beta_{20}$ and $\beta_{30}$: the collective potential, the three mass parameters $B_{22}$, $B_{23}$, $B_{33}$, 
and the moment of inertia $\mathcal{I}$. These functions are determined by constrained self-consistent mean-field calculations 
for a specific choice of the nuclear energy density functional and pairing interaction. In the present implementation of the model the 
single-nucleon wave functions, energies, 
and occupation factors, generated from constrained self-consistent solutions of the relativistic mean-field  
plus BCS-pairing equations (RMF+BCS), provide the microscopic input for the parameters of the collective Hamiltonian.
The three mass parameters associated with the quadrupole and octupole collective coordinates are calculated in the 
perturbative cranking approximation, while the Inglis-Belyaev formula is used for the rotational moment of inertia~\cite{Xia2017_PRC96-054303}.  
From the diagonalization of the collective Hamiltonian one obtains the energy spectrum and the corresponding eigenfunctions 
that are used to calculate various observables, such as reduced transition probabilities. 

The dynamics of nuclear fission in the TDGCM+GOA approach 
is described by a local, time-dependent Schr\"odinger-like equation 
in the space of collective coordinates $\bm{q}$,
\begin{equation}
i\hbar \frac{\partial g(\bm{q},t)}{\partial t} = \hat{H}_{\rm coll} (\bm{q}) g(\bm{q},t) , 
\label{eq:TDGCM}
\end{equation}
where $g(\bm{q},t)$ is the complex wave function of the collective variables $\bm{q}$. 
In the present study of fission dynamics
we consider the two-dimensional (2D) collective space of deformation parameters 
$\beta_{20}$ and $\beta_{30}$, 
and omit the rotational collective degrees of freedom.   
The Hamiltonian $\hat{H}_{\rm coll} (\bm{q})$ of Eq.(\ref{eq:Hcoll}) is thus simplified to the form:
\begin{equation}
\hat{H}_{\rm coll} (\bm{q}) = - {\hbar^2 \over 2} \sum_{ij} {\partial \over \partial q_i} B^{-1}_{ij}(\bm{q}) {\partial \over \partial q_j} + V(\bm{q}).
\label{eq:Hcoll2}
\end{equation}
In the TDGCM+GOA nuclear fission is considered as an adiabatic process, while non-adiabatic effects arising from the coupling between 
collective and intrinsic degrees of freedom are not taken into account. 
The collective space is divided into an inner region with a single nuclear density distribution, 
and an external region that contains the two fission fragments. 
The set of scission configurations defines the hyper-surface that separates the two regions.  
The flux of the probability current through this
hyper-surface provides a measure of the probability of observing a given pair of fragments at time $t$.
Each infinitesimal surface element is associated with a given pair of fragments $(A_L, A_H)$, where $A_L$ and $A_H$ denote the 
lighter and heavier fragments, respectively.
The integrated flux $F(\xi,t)$ for a given surface element $\xi$ is defined as \cite{Regnier2018_CPC225-180}
\begin{equation}
F(\xi,t) = \int_{t_0}^{t} \int_{\xi} \bm{J}(\bm{q},t) \cdot d\bm{S}, 
\label{eq:flux}
\end{equation}
where $\bm{J}(\bm{q},t)$ is the current
\begin{equation}
\bm{J}(\bm{q},t) = {\hbar \over 2i} \bm{B}^{-1}(\bm{q}) [g^{*}(\bm{q},t) \nabla g(\bm{q},t) - g(\bm{q},t) \nabla g^{*}(\bm{q},t)].
\end{equation}
The yield for the fission fragment with mass $A$ is defined by 
\begin{equation}
Y(A) \propto \sum_{\xi \in \mathcal{A}} \lim_{t \rightarrow \infty} F(\xi,t).
\end{equation}
The set $\mathcal{A}(\xi)$ contains all elements belonging to the scission hyper-surface such that one of the fragments has mass number $A$.

To describe the dynamics of induced fission, we assume that the compound nucleus is in a state of thermal equilibrium at temperature $T$, 
and the potential entering the collective Hamiltonian Eq.~(\ref{eq:Hcoll2}) is given by the Helmholtz free energy 
$F=E(T)-TS$, with $E(T)$ the mean-field (RMF+BCS) deformation energy in the $(\beta_{20},\beta_{30})$ plane, and $S$ is the 
entropy of the compound system. The mass tensor is calculated in the finite-temperature perturbative cranking 
approximation~\cite{Zhu2016_PRC94-024329,Martin2009_IJMPE18-861,Zhao2019_PRC99-014618}.
The initial collective wave packet $g(\bm{q},t=0)$ is constructed as described in Ref.~\cite{Zhao2019_PRC99-014618}, 
and the average energy of the collective initial state $E_{\rm coll.}^{*}$ is chosen to be 1 MeV 
above the highest fission barrier. 

The collective potential (Helmholtz free energy at finite temperature) and the mass tensor are determined by 
microscopic self-consistent mean-field calculations based on universal energy density functionals (EDFs). 
We employ the point-coupling relativistic EDF DD-PC1~\cite{Niksic2008_PRC78-034318}.  
Pairing correlations are taken into account in the BCS approximation and here, 
as in Ref.~\cite{Zhao2015_PRC92-064315}, we use a separable
pairing force of finite range:
\begin{equation}
V(\mathbf{r}_1,\mathbf{r}_2,\mathbf{r}_1^\prime,\mathbf{r}_2^\prime) = G_0 ~\delta(\mathbf{R}-
\mathbf{R}^\prime) P (\mathbf{r}) P(\mathbf{r}^\prime) \frac{1}{2} \left(1-P^\sigma\right),
\label{pairing}
\end{equation}
where $\mathbf{R} = (\mathbf{r}_1+\mathbf{r}_2)/2$ and $\mathbf{r}=\mathbf{r}_1- \mathbf{r}_2$
denote the center-of-mass and the relative coordinates, respectively. $P(\mathbf{r})$ reads 
\begin{equation}
P(\mathbf{r})=\frac{1}{\left(4\pi a^2\right)^{3/2}} e^{-\mathbf{r}^2/4a^2}.
\end{equation}
The two parameters of the interaction were originally 
adjusted to reproduce the density dependence of the pairing gap in nuclear matter at the
Fermi surface computed with the D1S parameterization of the Gogny force~\cite{Berger1991_CPC63-365}.

The deformation-dependent energy surface is obtained in a self-consistent finite-temperature mean-field 
calculation with constraints on the mass multipole moments $Q_{\lambda\mu} = r^\lambda Y_{\lambda \mu}$.
The nuclear shape is parameterized by the deformation parameters
\begin{equation}
 \beta_{\lambda\mu} = {4\pi \over 3AR^\lambda} \langle Q_{\lambda\mu} \rangle.
\end{equation}
The shape is assumed to be invariant under the exchange of the $x$ and $y$ axes, 
and all deformation parameters $\beta_{\lambda\mu}$ with even $\mu$ can be included simultaneously.
The self-consistent RMF+BCS equations are solved by an expansion in the 
axially deformed harmonic oscillator (ADHO) basis~\cite{Gambhir1990_APNY198-132}.
In the present study calculations have been performed 
in an ADHO basis truncated to $N_f = 20$ oscillator shells.
For the details of the multidimensionally constrained RMF+BCS model we refer the 
reader to Refs.~\cite{Lu2014_PRC89-014323,Zhao2019_PRC99-014618}.

\section{\label{sec:results}From ground-state deformation to the formation of fission fragments: results and discussion}

The present study starts with an analysis of collective spectra and induced fission dynamics of $^{228}$Th, that 
illustrates the capability of the GCM+GOA approach to describe both static and dynamic aspects of nuclear structure governed by 
collective degrees of freedom. The collective coordinates are the axially symmetric quadrupole 
 $\beta_{20}$ and octupole $\beta_{30}$ deformation parameters. To obtain the eigenspectrum of the collective Hamiltonian 
 we have performed a deformation-constrained zero-temperature self-consistent RMF+BCS calculation of the potential energy surface 
and single-nucleon wave functions. To reproduce the empirical pairing gaps in this mass region, the strength parameters 
of the pairing force have been increased 
with respect to the original values by the following factors $G_{n}/G_{0}=1.12$ and $G_{p}/G_{0}=1.08$.
The self-consistent solutions determine the parameters of the collective Hamiltonian Eq. (\ref{eq:Hcoll}). 

The analysis of induced fission dynamics is based on the corresponding 
self-consistent finite-temperature RMF+BCS calculation that produces a deformation energy surface $F(\bm{q})$, 
and variations of the free energy between two points $\bm{q}_1$ and $\bm{q}_2$ are given by  
$\delta F |_{T} = F(\bm{q}_{1}, T) - F(\bm{q}_{2}, T)$~\cite{Schunck2015_PRC91-034327}. 
The internal excitation energy $E_{\rm int.}^{*}$ of a nucleus at temperature $T$ is defined 
as the difference between the total binding energies of the equilibrium 
RMF+BCS minimum at temperature $T$ and at $T=0$. The time evolution of the initial GCM+GOA wave packet, governed by the 
collective Hamiltonian Eq. (\ref{eq:Hcoll2}), is computed with the TDGCM+GOA computer 
code FELIX (version 2.0)~\cite{Regnier2018_CPC225-180}.
The time step is $\delta t=5\times 10^{-4}$ zs (1 zs $= 10^{-21}$ s), and the charge and mass 
distributions are calculated after $10^{5}$ time steps, which correspond to 50 zs.
The scission configurations are defined by using the Gaussian neck operator $\hat{Q}_{N}=\exp[-(z-z_{N})^{2} / a_{N}^{2}]$, 
where $a_{N}=1$ fm and $z_{N}$ is the position of the neck~\cite{Younes2009_PRC80-054313}.
We define the pre-scission domain by $\langle \hat{Q}_{N} \rangle>3$, and consider the frontier of this domain as the scission contour.
Just as in our previous studies of Refs.~\cite{Tao2017_PRC96-024319,Zhao2019_PRC99-014618}, 
the parameters of the additional imaginary absorption potential that takes into account the escape 
of the collective wave packet  in the domain outside the region of calculation \cite{Regnier2018_CPC225-180} are: 
the absorption rate $r=20\times 10^{22}$ s$^{-1}$, and the width of the absorption band $w=1.5$.
Following Refs.~\cite{Regnier2016_PRC93-054611,Zhao2019_PRC99-014618}, 
the fission yields are obtained by convoluting the raw flux with a Gaussian function of the number of particles. 
The width is set to 1.6 units for the charge yields.

\subsection{\label{subsec:spectrum}Collective excitation spectrum of $^{228}$Th}

\begin{figure}
 \includegraphics[width=0.48\textwidth]{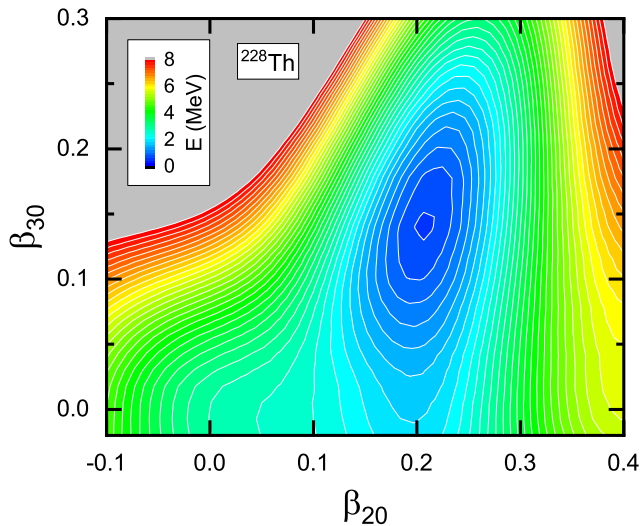}
\caption{(Color online)~\label{fig:Th228_Coll}%
Axially-symmetric quadrupole-octupole energy surface in the 
$\beta_{20} - \beta_{30}$ plane for $^{228}$Th.
The contours join points on the surface with the same energy, 
and the separation between neighboring contours is 0.2 MeV.
}
\end{figure}

\begin{figure}
 \includegraphics[width=0.48\textwidth]{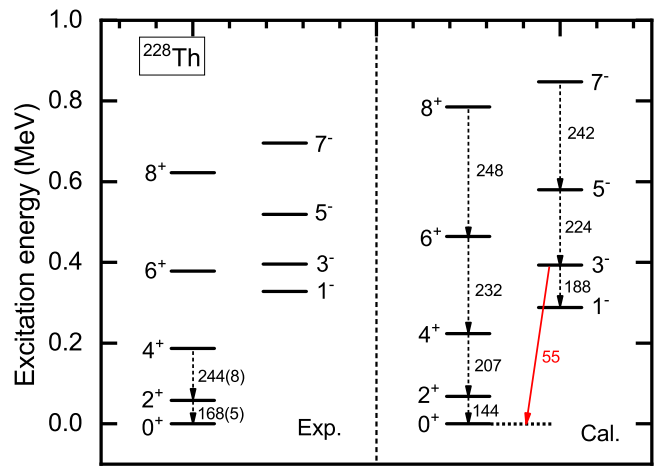}
\caption{(Color online)~\label{fig:Th228_Level}%
Experimental~\cite{NNDC,Neal1965_PR137-B1164} and calculated 
yrast states of positive and negative parity in $^{228}$Th. 
The in-band $B({\rm E2})$ values (dotted) and the $B({\rm E3};3_{1}^{-} \rightarrow 0_{1}^{+})$ (solid) 
(both in Weisskopf units) are also shown.
}
\end{figure}

In the theoretical framework based on relativistic energy density functionals, 
the evolution of quadrupole and octupole shapes in 
thorium isotopes has been explored and successfully described 
using the collective Hamiltonian QOCH \cite{Li2013_PLB726-866,Li2016_JPG43-024005}, 
and the interacting boson model (IBM) \cite{Nomura2013_PRC88-021303R,Nomura2014_PRC89-024312}. 
Figure \ref{fig:Th228_Coll} displays the contour plot in the ($\beta_{20}, \beta_{30}$) 
plane of the deformation energy surface of $^{228}$Th in the region around the equilibrium minimum, 
obtained at zero temperature by imposing constraints on the expectation values of the mass 
quadrupole moment $\langle \hat{Q}_{20}\rangle$  
and octupole moment $\langle \hat{Q}_{30}\rangle$.
The plots are symmetric with respect to the $\beta_{30}$ axis. 
The energy surface exhibits a global minimum at $(\beta_{20}, \beta_{30}) \approx (0.2, 0.15)$, 
and it is rather soft along the octupole direction. Similar topologies have also been predicted by 
earlier self-consistent mean-field calculations, based on both non-relativistic~\cite{Robledo2011_PRC84-054302} 
and relativistic energy density functionals~\cite{Nomura2013_PRC88-021303R,Nomura2014_PRC89-024312,
Li2013_PLB726-866,Li2016_JPG43-024005,Agbemava2016_PRC93-044304}. 
The single-nucleon wave functions, energies, and occupation factors, 
determine the parameters of the QOCH as described in Sec.~\ref{sec:model}.
The resulting low-energy spectrum of collective positive-parity and negative-parity yrast states of $^{228}$Th, 
including the intraband $B(E2)$ values and the $B(E3; 3_{1}^{-} \rightarrow 0_{1}^{+})$ value
(both in Weisskopf units) are plotted in Fig.~\ref{fig:Th228_Level}, and compared with available 
data \cite{NNDC,Neal1965_PR137-B1164}. For the excitation energies a very good agreement with experiment is obtained, 
except for the fact that the empirical moment of inertia is larger than that predicted by the
collective Hamiltonian. This is a well known effect of using the simple Inglis-Belyaev 
approximation for the moment of inertia. The wave functions, however, are not 
affected by this approximation and we note that the model 
reproduces the intraband $E2$ transition probabilities.
The negative-parity band is located close in energy to the ground-state positive-parity band, 
and its low excitation energy reflects the degree of octupole correlations in the equilibrium 
minimum, as well as the softness of the potential in the $\beta_{30}$ direction.

\subsection{\label{subsec:yields}Induced fission: charge fragment distributions}

\begin{figure}
 \includegraphics[width=0.48\textwidth]{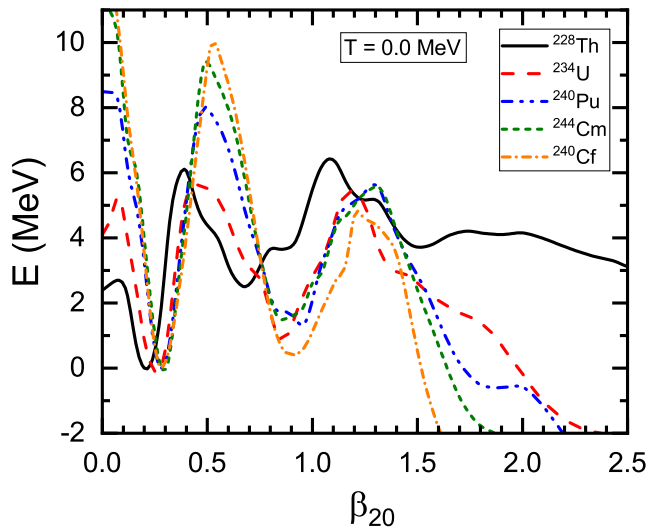}
\caption{(Color online)~\label{fig:Th_Cf_1DPES}%
Deformation energy curves (in MeV) along the least-energy fission path as 
functions of the quadrupole deformation parameter $\beta_{20}$, for 
$^{228}$Th, $^{234}$U, $^{240}$Pu, $^{244}$Cm, and $^{250}$Cf.
All curves are normalized to their values at equilibrium minimum.
}
\end{figure}

\begin{figure*}
 \includegraphics[width=0.70\textwidth]{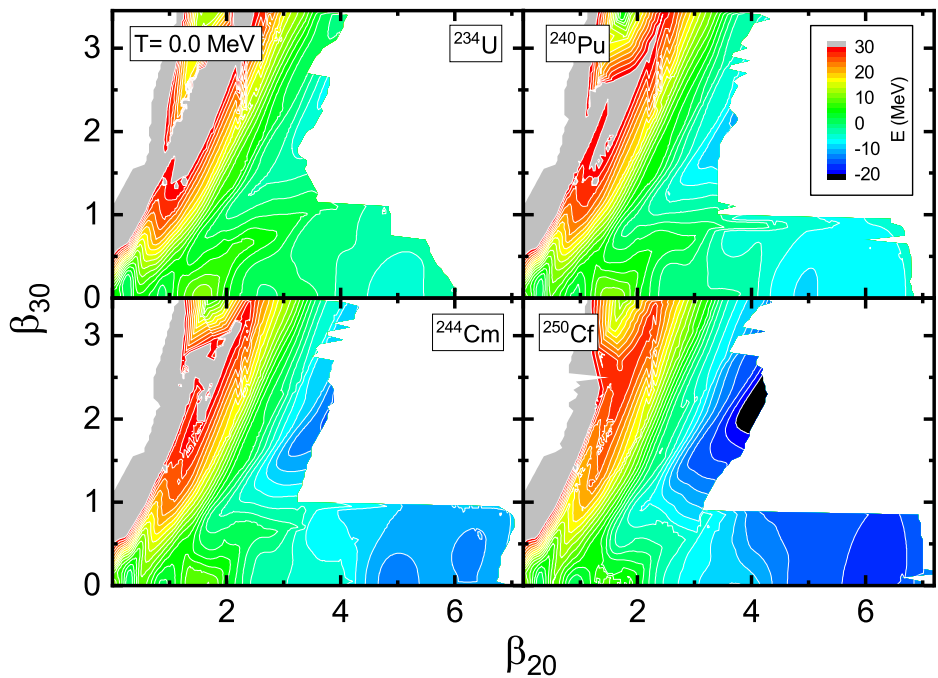}
\caption{(Color online)~\label{fig:U_Cf_2DPES}%
Axially-symmetric quadrupole-octupole deformation energy surfaces in the 
($\beta_{20}, \beta_{30}$) plane for $^{234}$U, $^{240}$Pu, $^{244}$Cm, and $^{250}$Cf.
In each panel the energies are normalized with respect to the corresponding value at the equilibrium minimum.
The contours join points on the surface with the same energy  
and the separation between neighboring contours is 2.0 MeV.
}
\end{figure*}

\begin{figure}
 \includegraphics[width=0.48\textwidth]{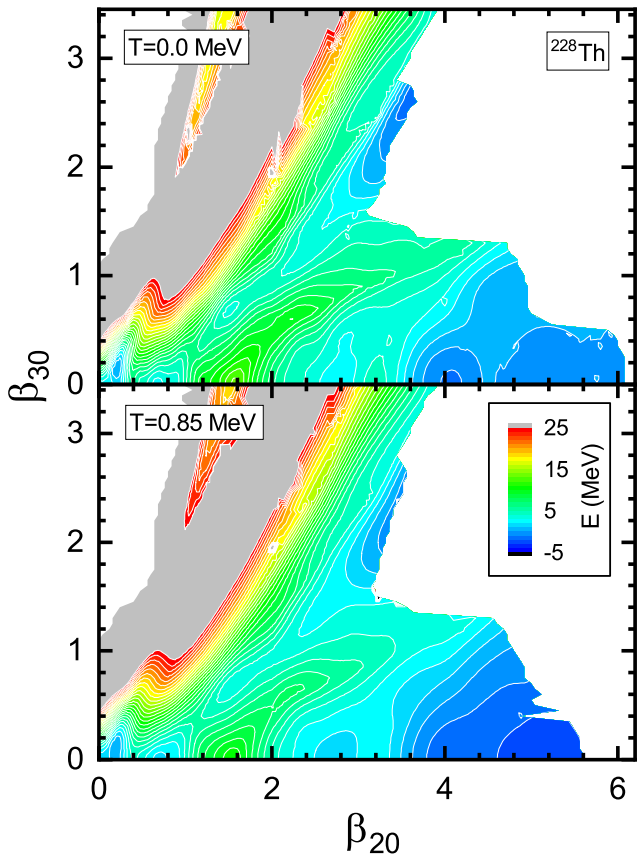}
\caption{(Color online)~\label{fig:Th228_FT_PES}%
Deformation (free) energy (in MeV) of $^{228}$Th in the ($\beta_{20}, \beta_{30}$) plane 
at zero temperature and at $T=0.85$ MeV.
In both panels the energies are normalized with respect to the corresponding value 
at the equilibrium minimum, and the contours join points on the surface with the same energy.
The contour interval is 1.0 MeV.
}
\end{figure}

\begin{figure}
 \includegraphics[width=0.48\textwidth]{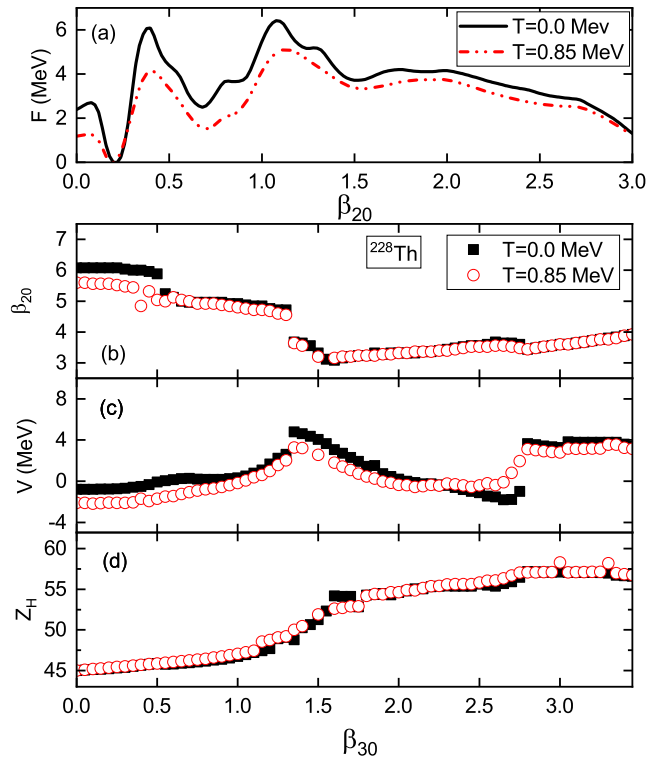}
\caption{(Color online)~\label{fig:Th228_Sec}%
Free energy (in MeV) of $^{228}$Th along the least-energy fission path 
as function of the quadrupole deformation (a).  
The values of the deformation parameter $\beta_{20}$ (b), the free energy (c),
and the heavy-fragment charge number, along the frontier of the domain 
defined by $Q_{N} > 3.0$ at zero temperature and at  
$T=0.85$ MeV. The position on the scission contour is labeled by the corresponding $\beta_{30}$ value.
}
\end{figure}

\begin{figure}
\includegraphics[width=0.48\textwidth]{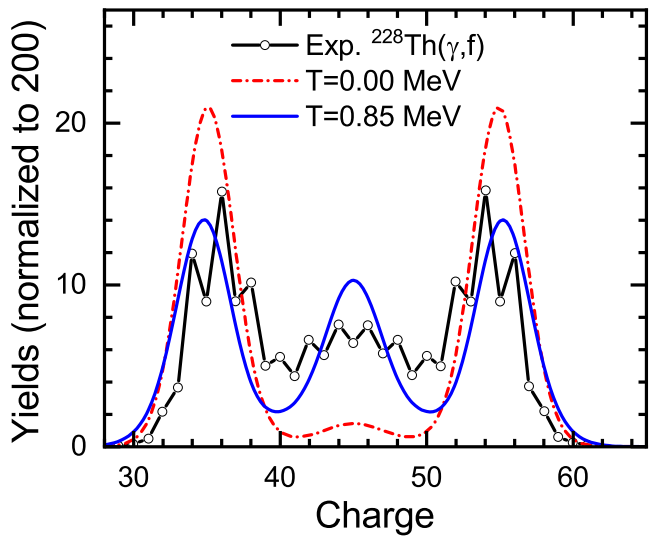}
\caption{(Color online)~\label{fig:ChargeYields_Th}%
Charge yields for photo-induced fission of $^{228}$Th. 
The collective potentials and perturbative cranking inertia tensors for zero and  
 finite temperature are used in the TDGCM+GOA calculation. $T=0.85$ MeV corresponds to 
 the intrinsic excitation energy of $E_{\rm int}^{*} \approx 11$ MeV, equivalent to the 
 peak value of the photon energy distribution. 
}
\end{figure}

\begin{figure*}
\includegraphics[width=0.70\textwidth]{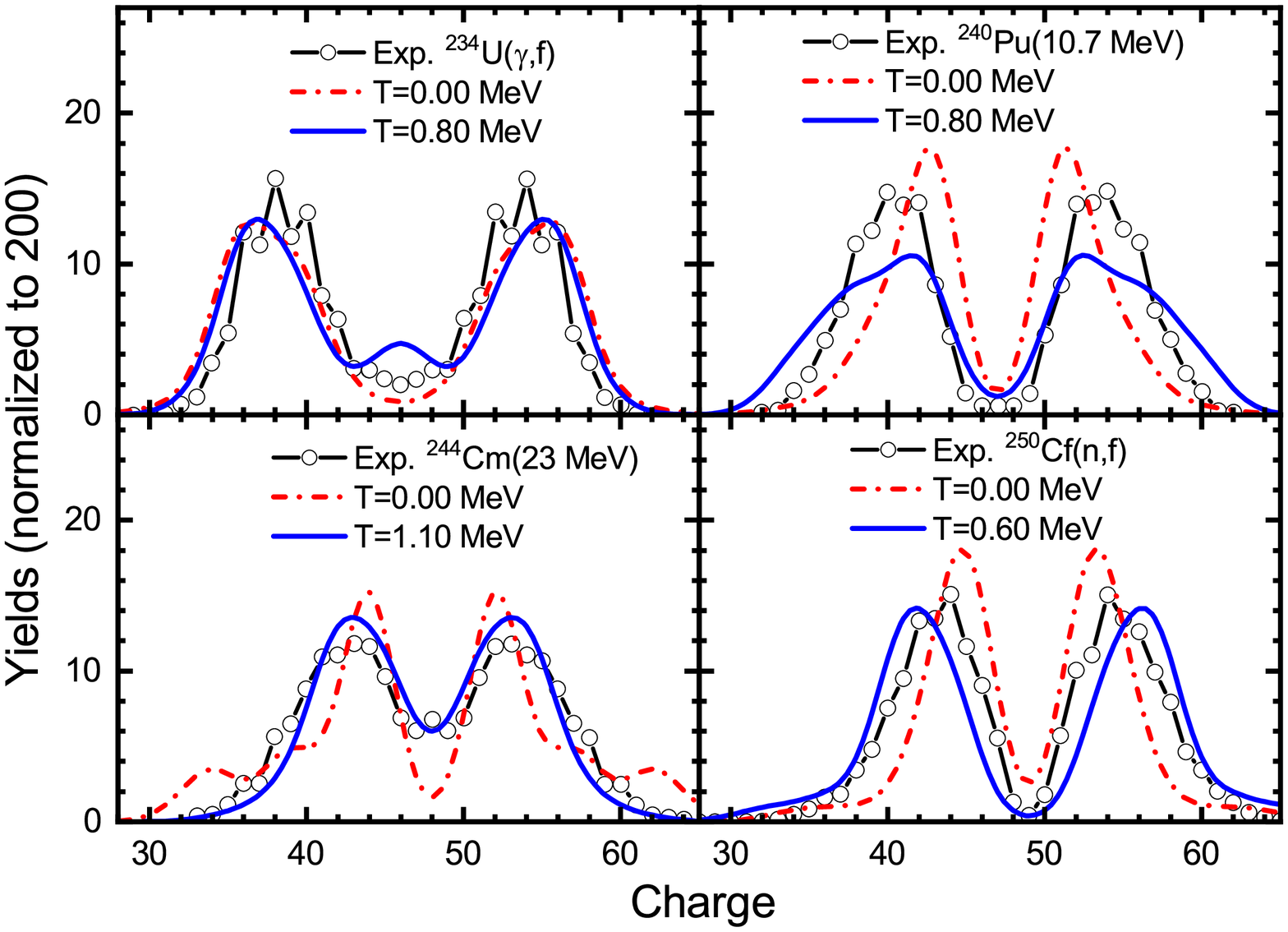}
\caption{(Color online)~\label{fig:ChargeYields}%
Same as in Fig.~\ref{fig:ChargeYields_Th}, but for $^{234}$U, $^{240}$Pu, $^{244}$Cm, and $^{250}$Cf.
See text for the description.}
\end{figure*}


In Fig.~\ref{fig:Th_Cf_1DPES} we plot the deformation energy curves as functions of 
the quadrupole deformation parameter $\beta_{20}$, along the least-energy fission paths 
of $^{228}$Th, $^{234}$U, $^{240}$Pu, $^{244}$Cm, and $^{250}$Cf at zero temperature.
A triple-humped barrier is predicted along the static fission path for $^{228}$Th,
with the barrier heights $6.06$, $6.42$, and $4.20$ MeV from the inner to the outer barrier, respectively. 
This is consistent with previous results obtained in Ref.~\cite{Zhao2015_PRC91-014321} by using the energy 
density functionals DD-ME2~\cite{Lalazissis2005_PRC71-024312} and PC-PK1~\cite{Zhao2010_PRC82-054319}.
Two-humped barriers are calculated for the other four nuclei, and 
the inner barrier heights are: $5.62$, $8.09$, $9.25$, and $9.97$ MeV
for $^{234}$U, $^{240}$Pu, $^{244}$Cm, and $^{250}$Cf, respectively.
The heights of the outer barriers are very similar for these nuclei: 
$5.41$, $5.61$, $5.54$, and $4.69$ MeV, respectively.
The corresponding $\beta_{20}$-$\beta_{30}$ deformation energy surfaces 
are shown in Figs.~\ref{fig:U_Cf_2DPES} and \ref{fig:Th228_FT_PES}.
Only the points which correspond to self-consistent solutions with $\langle \hat{Q}_{N} \rangle \geq 3$ are plotted, 
and the frontier of this domain determines the scission contour.
The topography of the quadrupole-octupole energy surfaces are similar for these nuclei, 
and one notices the ridge separating the asymmetric and symmetric fission valleys.  

The evolution of deformation energy surfaces and barrier heights with temperature 
has been discussed in detail in our previous study of finite-temperature effects on 
fission dynamics \cite{Zhao2019_PRC99-014618}. Here, in particular, we display 
in Fig.~\ref{fig:Th228_FT_PES} the quadrupole-octupole free energy surface of $^{228}$Th 
at zero temperature and at $T=0.85$ MeV.
The corresponding internal excitation energy $E_{\rm int}^{*}$ is approximately 11 MeV.
The self-consistent zero temperature and $T=0.85$ MeV free energy surfaces are similar, 
but the ridge separating the asymmetric and symmetric fission valleys decreases with temperature.
The free energy along the asymmetric least-energy fission path at $T=0$ and $0.85$ MeV
are compared in Fig.~\ref{fig:Th228_Sec} (a). We notice that 
the barriers are considerably lowered at $T=0.85$ MeV, especially the inner two. 
From the inner to the outer barrier, the heights are: $4.15$, $5.11$, and $3.75$ MeV. 
The scission contour at $T=0$ and $0.85$ MeV displays similar patterns.
It starts from an elongated symmetric point at $\beta_{20}>5.5$, and evolves to a minimal elongation
$\beta_{20} \sim 3$ as asymmetry increases.  
The values of the quadrupole deformation $\beta_{20}$, the free energy, and the 
heavy fragment charge numbers along the scission contour  
are plotted as functions of $\beta_{30}$ in Fig.~\ref{fig:Th228_Sec} (b), (c), and (d).
For these quantities the differences between zero-temperature and $T=0.85$ MeV 
along the scission contour are indeed very small. 

The dynamics of induced fission of $^{228}$Th, $^{234}$U, $^{240}$Pu, $^{244}$Cm, and $^{250}$Cf 
is explored by following the time evolution of an initial wave packet $g(\bm{q},t=0)$, built 
as a Gaussian superposition of the quasi-bound states $g_k$, 
\begin{equation}
g(\bm{q},t=0) = \sum_{k} \exp\left( { (E_k - \bar{E} )^{2} \over 2\sigma^{2} } \right) g_{k}(\bm{q}),
\label{eq:initial-state}
\end{equation}
where the value of the parameter $\sigma$ is set to 0.5 MeV. The collective states $\{ g_{k}(\bm{q}) \}$ 
are solutions of the stationary eigenvalue equation in which the original collective potential $V(\bm{q})$ is replaced by a 
new potential $V^{\prime} (\bm{q})$ that is obtained by extrapolating the inner potential barrier with a quadratic form. 
The mean energy $\bar{E}$ in Eq.~(\ref{eq:initial-state}) is then adjusted iteratively in 
such a way that $\langle g(t=0)| \hat{H}_{\rm coll} | g(t=0) \rangle = E_{\rm coll.}^{*}$. The 
TDGCM+GOA Hamiltonian of Eq.~(\ref{eq:Hcoll2}), with the original collective potential 
$V(\bm{q})$, propagates the initial wave packet in time (cf. Eq.~(\ref{eq:TDGCM})).
For finite-temperature calculations the temperature is chosen in such a way that the internal excitation energy 
$E_{\rm int}^{*}$ corresponds to the experimental excitation energy of the compound nucleus.
At finite temperature the collective potential corresponds to the Helmholtz free energy 
$F=E(T)-TS$, with $E(T)$ the RMF+BCS deformation energy, 
and the mass tensor is calculated using the finite-temperature perturbative cranking approximation. At each point 
of the deformation energy surface the entropy of the compound nuclear system is computed using the self-consistent 
thermal occupation probabilities of single-quasiparticle states \cite{Zhao2019_PRC99-014618}.

The charge yields obtained with the TDGCM+GOA, and normalized to $\sum_{A} Y(A) = 200$, 
are shown in Figs.~\ref{fig:ChargeYields_Th} and \ref{fig:ChargeYields}, 
in comparison to the experimental fragment charge distributions.
For $^{228}$Th, already the calculation at zero temperature reproduces the trend of 
the data except, of course, the odd-even staggering. 
The predicted asymmetric peaks are located at $Z=35$ and $Z=55$, 
one mass unit away from the experimental peaks at $Z=36$ and $Z=54$~\cite{Schmidt2000_NPA665-221}.
The asymmetric yields are overestimated, while the symmetric yields are markedly underestimated. 
The data for $^{228}$Th correspond to photo-induced fission with photon energies in the interval $8-14$ MeV, and a 
peak value of $E_{\gamma} = 11$ MeV \cite{Schmidt2000_NPA665-221}. 
The charge yields obtained at temperature $T=0.85$ MeV, which corresponds to the 
intrinsic excitation energy $E_{\rm int}^{*} \approx 11$ MeV, 
are also shown in Fig.~\ref{fig:ChargeYields_Th} (solid curve). We notice that,
by including the finite-temperature effect, the predicted asymmetric yields are lowered 
and the symmetric yields are enhanced, producing a much better agreement with the experiment.
For $^{234}$U the data were also obtained in photo-induced fission 
with $E_{\gamma} = 11$ MeV for the peak photon energy~\cite{Schmidt2000_NPA665-221}.
The corresponding temperature is $T=0.80$ MeV. 
The charge yields obtained at $T=0$ and $0.80$ MeV are very similar, though one finds 
a small enhancement of the symmetric yield due to finite-temperature effect.

The experimental charge yields of $^{240}$Pu and $^{244}$Cm are taken from Ref.~\cite{Ramos2018_PRC97-054612}.
The average excitation energies are $10.7$ and $23$ MeV, and correspond to the temperatures 
$T=0.80$ and $1.10$ MeV for $^{240}$Pu and $^{244}$Cm, respectively.
For both nuclei the charge yields exhibit a two-humped structure, and our 
calculation clearly reproduces the trend of the data.
For $^{240}$Pu the charge yields at zero temperature overestimate the experimental asymmetric peaks,  
and the calculated peaks do not agree quantitatively with the experimental locations. A much 
better agreement is obtained by considering the finite temperature of the compound nucleus, even though the 
experimental asymmetric yields are somewhat underestimated by the calculation at $T=0.80$ MeV.
Perhaps the strongest finite-temperature effect is found for $^{244}$Cm, 
for which the calculated distribution of charge yields at zero temperature differs considerably from the experimental 
results. 
This is, of course, due to the fact that the data correspond to a rather high excitation energy of $23$ MeV. 
One therefore expects that the deformation energy surface and the inertia tensor at the corresponding 
temperature  $T=1.1$ MeV will markedly differ from those obtained at zero temperature. 
In fact, the predicted charge yields at $T=1.1$ MeV for the compound system are in excellent agreement 
with the data, and reproduce both the shape of the empirical distribution, as well as the yields and location of 
the peaks. 

Finally, in the case of $^{250}$Cf the charge yields distribution obtained at zero temperature overestimates both the asymmetric peak yields and the symmetric yields, and does not reproduce the empirical width and location of the 
peaks resulting from thermal neutron induced fission.
We have thus calculated the charge yields distribution at $T=0.6$ MeV, which is consistent with the 
experimental excitation energy. The inclusion of finite-temperature effect produces a lowering of the 
asymmetric peaks and symmetric charge yields, leading to a much
improved agreement with the data \cite{Brown2018_NDS148-1}. 
We note that in all cases investigated in the present study the predicted heavy fission fragments exhibit peaks  
between $Z=52$ and $Z=56$, in 
excellent agreement with the TDDFT results of Ref.~\cite{Scamps2018_Nature564-382} and with experiment.
\bigskip

\section{\label{sec:summary}Summary}

Using the microscopic TDGCM+GOA framework based on the relativistic energy density functional DD-PC1 and a separable 
pairing force of finite range, we have shown that it is possible to simultaneously describe collective excitation spectra 
of actinide nuclei in the octupole deformed equilibrium potential well, and the dynamics of the entire fission process 
in the two dimensional collective space of axial quadrupole and octupole deformations $(\beta_{20},\beta_{30})$. 

Our previous studies have shown that a GCM+GOA quadrupole-octupole collective Hamiltonian provides an accurate 
description of spectroscopic properties (low-energy positive- and negative-parity bands, average octupole deformations, and 
transition rates) of nuclei characterized by pronounced octupole mean-field deformations, both in the region of actinides 
that can undergo spontaneous or induced fission, and in the region of even-even 
medium-heavy nuclei ($54\leq Z\leq 64$) where the heavier fission fragments are found. In the present study this 
is illustrated with a brief analysis of low-energy yrast positive- and negative-parity states of $^{228}$Th. 

Starting from the initial Gaussian superposition of eigenstates of the collective Hamiltonian in the 
equilibrium potential well, with an average energy chosen $\approx 1$ MeV above the fission barrier, 
the TDGCM+GOA propagates the collective wave packet in time through the scission hyper-surface. 
The corresponding flux of the probability current determines the mass and charge fragment distributions. 
In addition to $^{228}$Th, we have also computed the charge yields for induced fission of $^{234}$U, $^{240}$Pu, 
$^{244}$Cm, and $^{250}$Cf. The calculation reproduces the trend of the data already at zero temperature, but 
in general the collective potential and mass parameters are affected by the increase of internal excitation energy 
in induced fission. Therefore, to describe the dynamics of induced fission we use a finite-temperature extension of 
nuclear density functional theory, and assume that the compound nucleus is in a state of thermal equilibrium at 
a temperature that corresponds to the internal excitation energy. In this approximation the collective 
potential corresponds to the Helmholtz free energy and the mass tensor is calculated using the finite-temperature 
perturbative cranking formula. Even though the model is still based on the adiabatic approximation, 
the extension of TDGCM+GOA to finite temperature leads to a considerable improvement of the 
calculated charge yields. In general, the theoretical yields are in very good agreement with available data and, 
in particular, the peaks of the charge distribution for the heavy fragments are predicted between 
$Z=52$ and $Z=56$. These results are consistent with the findings of the TDDFT study of Ref.~\cite{Scamps2018_Nature564-382}, 
in which the final charge asymmetry of the fragments has been 
attributed to the extra binding of the heavier fragments with shell-stabilized octupole deformations.  

\bigskip
\acknowledgements
This work has been supported by the Inter-Governmental S\&T Cooperation Project between China and Croatia. 
It has also been supported in part by the QuantiXLie Centre of Excellence, a project co-financed by the Croatian Government and European Union through the European Regional Development Fund - the Competitiveness and Cohesion Operational Programme (KK.01.1.1.01).
Calculations have been performed in part at the HPC Cluster of KLTP/ITP-CAS and the Supercomputing Center, CNIC of CAS. S.G.Z. was supported by the National Key R\&D Program of China (2018YFA0404402), the NSF of China (11525524, 11621131001, 11647601, 11747601, and 11711540016), the CAS Key Research Program of Frontier Sciences (QYZDB-SSWSYS013), the CAS Key Research Program (XDPB09), and the IAEA CRP ``F41033''. 
Z.P.L. was supported by the NSFC (11875225, 11790325).

\bibliographystyle{apsrev4-1}


%

\end{document}